%% file: main.tex
\documentclass{article}

\usepackage[final, nonatbib]{neurips_2022}

\usepackage[utf8]{inputenc} 
\usepackage[T1]{fontenc}    
\usepackage{hyperref}       
\usepackage{url}            
\usepackage{booktabs}       
\usepackage{amsfonts}       
\usepackage{nicefrac}       
\usepackage{microtype}      
\usepackage{xcolor}         

\usepackage{enumitem}
\usepackage[pdftex]{graphicx}
\usepackage{grffile}
\usepackage{amsmath}
\usepackage{amssymb}
\usepackage{amsfonts}
\usepackage[bb=dsserif]{mathalpha}
\usepackage{algorithm}
\usepackage{algorithmic}
\usepackage{colortbl}
\usepackage{multirow}
\usepackage{multicol}
\usepackage{wrapfig}
\usepackage{caption}

\graphicspath{{figures/}} 
\usepackage{acronym}
\acrodef{rle}[Ring-LWE]{Ring Learning With Errors}
\acrodef{FHE}[FHE]{Fully Homomorphic Encryption }
\acrodef{ntt}[NTT]{Number Theoretic Transform}
\acrodef{fft}[FFT]{Fast Fourier Transform}
\acrodef{dft}[DFT]{Discrete Fourier Transform}
\acrodef{fifo}[FIFO]{First-In First-Out}
\acrodef{rom}[ROM]{Read Only Memory}
\acrodef{ka}[Karatsuba]{Karatsuba}
\acrodef{pqc}[PQC]{Post-Quantum Cryptography}
\acrodef{ringeq}[$\mathbb{Z}_q$]{$\mathbb{Z}_q{\scriptstyle[\mkern1mu}x{\scriptstyle\mkern1mu]}{\scriptstyle <}x^n{\scriptstyle +}{\tiny 1}{\scriptstyle>}$}

\title{A Fully Pipelined FIFO Based Polynomial Multiplication Hardware Architecture Based On Number Theoretic Transform}

\author{%
  Moslem Heidarpur \\
  University of Windsor \\
  \tt{heidarp@uwindsor.ca} \\
  \And
  Mitra Mirhassani \\
  University of Windsor \\
  \tt{mitramir@uwindsor.ca} \\
  \And 
  Norman Chang \\
  Ansys\\
  \tt{norman.chang@ansys.com} \\
 }

\begin{document}

\maketitle

\setcounter{footnote}{0}
\input{sections/0_abstract}
\input{sections/1_intro}

\input{sections/2_background}

\input{sections/3_proposed}

\input{sections/4_results}

\input{sections/5_conclusion}

\begin{ack}
\vspace{-2mm}
Thank you
\end{ack}

{
\small
\bibliographystyle{plain}
\bibliography{reference.bib}
}

\clearpage

\appendix

\end{document}

%% file: sections/0_abstract.tex
\begin{abstract}
This paper presents digital hardware for computing polynomial multiplication using Number Theoretic Transform (NTT), specifically designed for implementation on Field Programmable Gate Arrays (FPGAs). Multiplying two large polynomials applies to many modern encryption schemes, including those based on Ring Learning with Error (RLWE). The proposed design uses First In, First Out (FIFO) buffers to make the design fully pipelined and capable of computing two $n$ degree polynomials in $n/2$ clock cycles. This hardware proposes a two-fold reduction in the processing time of polynomial multiplication compared to state-of-the-art enabling twice as much encryption in the same amount of time. Despite that, the proposed hardware utilizes fewer resources than the fastest-reported work.\footnote{Part of our code is publicly available at \tt\href{https://github.com/mheidarpur/fpntt}{https://github.com/mheidarpur/fpntt}}
\end{abstract}

%% file: sections/1_intro.tex
\section{Introduction}
\label{sec:intro}

Some of the widely used cryptography schemes such RSA \cite{rsa} , ECC \cite{roetteler2017quantum}, and AES \cite{aes} are based on the hardness of number theoretical problems that are very hard to solve on classical computers. However, quantum computers can solve these problems in polynomial time using the Shor \cite{shor} algorithm.
Since quantum computers are becoming a reality \cite{googleq,chinaq},
\ac{pqc} algorithms have gained greater attention. Lattice-based cryptography algorithms such \ac{rle} \cite{lwe}, and NTRU \cite{ntru}   are among the strong candidates for quantum-safe cryptography.
Lattice-based systems are based on the hardness of lattice problems and are believed to resist quantum computer attacks.

Furthermore, \ac{rle} is one of the algorithms that can be used for \ac{FHE} \cite{gentry} implementations. The encrypted data can be processed using \ac{FHE} without decrypting it first, which is useful in preserving privacy. Fully homomorphic encryption is not yet quite feasible, mainly due to the large number of computations required to perform homomorphic operations on the encrypted data.

Polynomial multiplication is the slowest and the most area-consuming operation in \ac{rle} cryptography systems. One efficient and fast algorithm widely used to perform polynomial multiplication is  \ac{ntt}   \cite{nttdu,ntt_longa, ntt_renter,ntt_feng,ntt_chen,ntt_mert, ntt_ulu, ntt_tan}.

An NTT polynomial multiplier hardware using a systolic array of the $n$-point NTT-core is presented in \cite{ntt_renter}.The work in \cite{nttdu} used various optimization algorithms to optimize the  \ac{ntt} based polynomial and reduce ROM usage. In \cite{ntt_feng}, authors proposed a fast polynomial multiplier using a multi-lane Stockham NTT algorithm. Chen et al. \cite{ntt_chen} propose an optimized  \ac{ntt} based polynomial multiplier by selecting an efficient parameter set and a pipelined architecture. Reference \cite{ntt_mert} proposed a  \ac{ntt} multiplier to be used as an accelerator for Encrypted Arithmetic Library (SEAL) \cite{seal}.

However, to be a suitable candidate for post-quantum cryptography, speed of operations is a very important factor. The works presented in \cite{nttdu, ntt_feng, ntt_chen, ntt_mert} are slow and therefore, cannot fulfil this need.
Additionally, improving the speed  of operation is vital for \ac{FHE} implementation,  considering the very large number of computations required. Reviewing the state-of-the-art, there is a gap in research with focus on improving the speed while at the same time having area-efficient  \ac{ntt} multipliers to be used in \ac{pqc} and  \ac{FHE} devices.

This work proposes a novel optimized hardware architecture to perform polynomial multiplication using  \ac{ntt}. The objective is to design a high speed polynomial multiplier while, at the same time, keeping the required resources low when compared with the state-of-art.

In order to accomplish this,  the order of performing operations was studied so that the pipelined hardware \ac{ntt}  could be completed with a minimal number of clock cycles. \ac{fifo} register shift buffers were used instead of memories to store data. Butterfly operations start at each stage of \ac{ntt} as soon as all required data become available.
Since the design is pipelined, it can perform $n$-point polynomial multiplication in $n/2$ number of clock cycles which is an increase by factor of two comparing to the fastest work reported.

The rest of the paper is organized as follows. Section II reviews the  \ac{ntt}, polynomial multiplication,  Karatsuba and Barret reduction algorithms. Section III discusses the architecture of the design.
Section IV  presents the details of the novel hardware presented for NTT-based pipelined polynomial multiplication. Section V presents the results and comparison with state-of-the-art. Finally, Section VI concludes the paper.

%% file: sections/2_background.tex
\vspace{-2.5mm}
\section{Background}
\label{sec:background}
\vspace{-1mm}
In Ring-LWE cryptography systems, operations are performed in the ring Rq. It is assumed that the ring is in the form of \acl{ringeq}, where $n$ is the polynomial degree, $q$ is the modulus of coefficients.
\begin{figure*}
    \centering
    \includegraphics[scale=0.5]{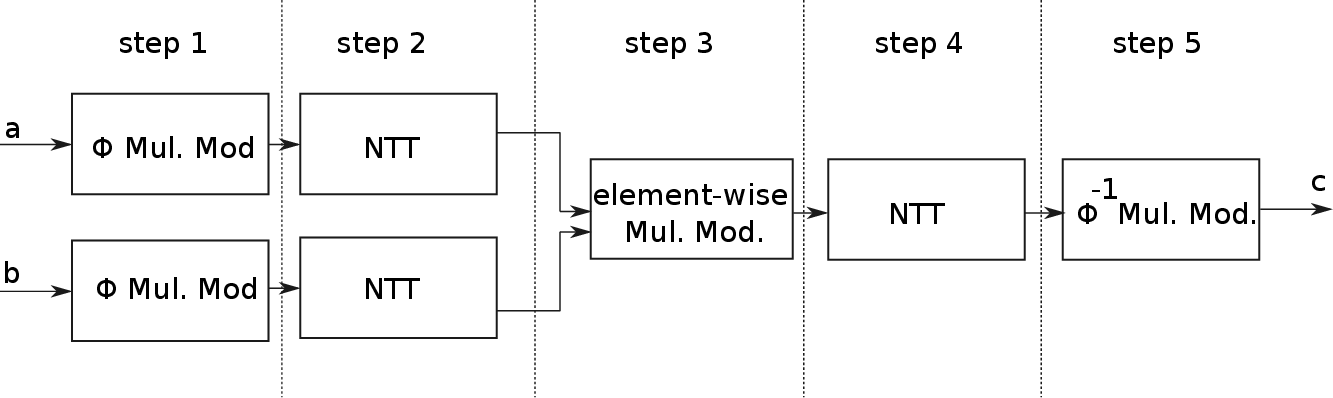}
    \caption{Polynomial multiplication using NTT and negative wrapped convolution could be divided into five steps. }
    \label{fig:structure}
\end{figure*}


Considering  $a(x)=\sum^{i=N-1}_{i=0} a_i x^i$ and $b(x)=\sum^{i=N-1}_{i=0} b_i x^i$ are two polynomial in the ring \acl{ringeq}, multiplication is performed as followed:
\begin{equation}
    c(x)=a(x) \times b(x)=\sum^{N-1}_{i=0}\sum^{N-1}_{j=0}a_ib_jx^{i+j}~mod~x^{N+1}
\end{equation}
The $c(x)$ has a complexity of $\mathcal{O}(N^2)$, which is high. The aim is to reduce the complexity using approaches such as \ac{ka} \cite{karats,overlap_free}, Toom-Cook \cite{toomcook}, or \ac{ntt} \cite{ntt_longa}.

The \ac{ntt}  is one of the common techniques to efficiently implement polynomial multiplications with the reduced complexity of $\mathcal{O}(Nlog(N))$.

\textbf{NTT Multiplication}:  For NTT operations, the value of $N$ should be a power of two, therefore, to multiply the two polynomials $a(x)$ and $b(x)$ are  padded with zeros to create $N$-point polynomials  $\tilde{a}(x)$ and  $\tilde{b}(x)$.

The NTT operation transforms an $N$-point polynomial $\tilde{a}(x)$ into a $N$-point NTT version, $\tilde{A}(x)$:
\begin{equation} 
    \tilde{A}(x)=NTT\left(\tilde{a}(x)\right)=\sum^{N-1}_{i=0} \tilde{A}_i x^i 
\end{equation}
The term $\tilde{A}_i$ represents the NTT coefficients and is defined as follows:
\begin{equation}
    \tilde{A}_i=\sum^{N-1}_{j=0}a_i\omega^{ij}
\end{equation}
The constant $\omega$, named the twiddle factor, should satisfy the condition $\omega^N \equiv 1~mod~M$, and $\omega^i\not\equiv 1~mod~M$ $\forall~ i<N$, where $M\equiv1~mod~N$

The multiplication output, $c(x)$, is then computed by taking the inverse NTT:
\begin{equation}
    c(x)=NTT^{-1}\left(  \tilde{A}(x).\tilde{B}(x) \right)
\end{equation}

However, to avoid the process of zero padding, a \ac{fft}-based weighted NTT can be used \cite{nconv}. 
This method, called negative wrapped convolution reduces the length of the NTT values by eliminating the need for zero padding.

\textbf{Fast
Fourier Transform (FFT)-based weighted NTT}: Multiplication using the negative wrapped convolution requires selecting a weighing coefficient ($\phi$). If $\phi \equiv 1 ~mod~2N$, then NTT multiplication result has  $mod~x^N+1$ does not need any polynomial reduction \cite{nconv}. The added cost is that coefficients need to be weighed before taking NTT and after performing the inverse NTT. Multiplication steps based on this approach, depicted in Fig. \ref{fig:structure},  are described as follows:

\begin{enumerate}
    \item Multiply each element of the polynomials by $\phi^i$ to construct weighted polynomials:
        \\ $\hat{a_i}= (a_i \times \phi^i) ~mod~ M$ \\
        $\hat{b_i}= (b_i \times \phi^i) ~mod~ M$
    \item Perform NTT on the weighted polynomials:
        \\
        $\hat{A}(x)=NTT(\hat{a}(x))$ \\
        $\hat{B}(x)=NTT(\hat{b}(x))$
    \item Element-wise multiplication of $\hat{A}$ and $\hat{B}$.\\
        $\bar{C}_i=(\hat{A_i} \times \hat{B_i})~mod~M$
    \item Take inverse NTT of $\bar{C}$\\
        $C=NTT^{-1}(\bar{C})$
    \item
        Scale and multiply each element of ${C}$ by $\phi^{-i}$ to obtain the multiplication result\\
        $c_i={C_i} \times \phi^{-i}$
\end{enumerate}

To optimize the critical path of the hardware, multiplication can be broken into smaller sub-multiplications. For example, at the lower level, \ac{ka} \cite{karats} can be used to perform integer multiplications.

\textbf{Karatsuba Multiplication}: To use the \ac{ka}, an $I$ with $l$ bit digit integer is considered as input. The polynomial version of $I$ can be written as follows:
\begin{equation}
    I(b)=i_S b+i_L
\end{equation}
where $i_S$ and $i_L$ are most significant and less significant bits and  $b=2^{l/2}$.

The \ac{ka} multiplication between two polynomials, $ I_1(b)$ and $I_2(b)$, could be performed as:
\begin{align}
    \label{ca2}
    I_1(b) \times I_2(b)&=(i_{1S} b+i_{1LS})(i_{2S} b+ i_{2L}) \nonumber \\
    & =i_{1S}.i_{2S}~b^2+ [(i_{1S}+i_{1L})  (i_{2S}+i_{2L}) \nonumber \\ &- i_{1L}.i_{2L} -i_{1S}.i_{1S}  ]~b + i_{1L}.i_{1L}
\end{align}
The multiplication in Eq. (\ref{ca2}) requires three sub-multiplications with $l/2$ size. Barret reduction \cite{bred} is used to preform the integer modulus operations.

\textbf{Barret Reduction}: Reduced integer $I_r$  can be calculated as:
\begin{align}
    \beta&= I~{\scriptstyle //}~M  \\
    I_r &= I - \beta I
\end{align}
where $\scriptstyle//$ is integer division, $\beta$ is division quotient, $I_r$ is reduced value of and is $\textstyle I_r~\equiv~I~mod~M$.
\begin{figure}
    \centering
    \includegraphics[scale=0.24]{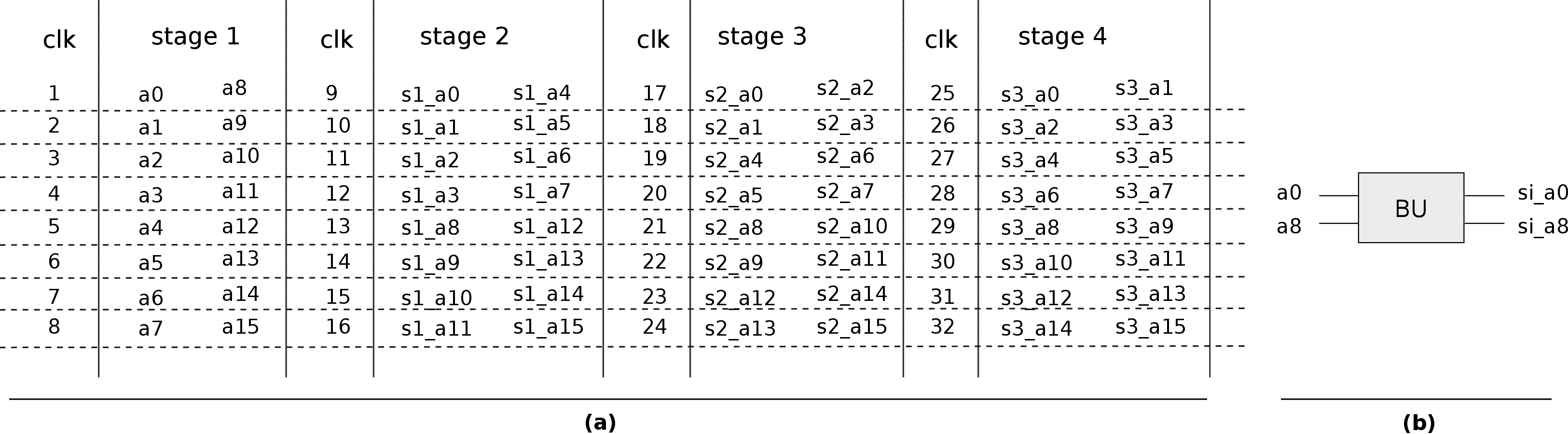}
    \caption{ (a) Performing naive NTT implementation over the polynomial $a=(a_{15} a_{14} ... a_1 a_0)$.  At clock one at stage 1, coefficients $a_0$ and $a_8$ are read from memory, butterfly operation is performed, and the result is stored in memory. (b) $si\_ak$ indicates the result of butterfly operation  on polynomial coefficients $ak$ in stage $si$.  Stage 1 takes eight clock cycles to complete. After this stage is completed, stage 2 performs butterfly on $s1\_a0$ and $s1\_a_4$. Computing NTT needs a total of 32 clock cycles to complete.  }
    \label{fig:ntt_normal}
\end{figure}

The process to perform reduction is described in the following.
\begin{enumerate}
    \item Integer $u$ is defined as $u =\lfloor  {\scriptstyle  \dfrac{2^k}{M}} \rfloor $, where $k$ is an integer, initially chosen to satisfy $2^k > M$ condition. \\
    \item Quotient $\beta$ can be calculated as:\\
        \begin{equation}
            \label{eq:barret_find_quotient}
            \beta= (Iu){\small >>}k
        \end{equation}
        where {\small $>> k$} symbolizes the shift to the right $k$ times.
    \item The remainder is:
        \begin{equation}
            \label{eq:barret_find_remainder}
            I_r = I - \lfloor \beta  \rfloor M
        \end{equation}
        where $\lfloor \beta  \rfloor$ is floor of $\beta$.
    \item  $I_r$ would be  the range of $[0,2M)$ since  $\dfrac{1}{M}$ $\leq$ $\dfrac{u}{2^k}$, therefore if ( $I_r \geq M$ ) \\
        \begin{equation}
            \label{eq:barret_find_final_remainder}
            I_r = I_r-M
        \end{equation}
\end{enumerate}
To find integer  k, error is $e$ was calculated:
\begin{equation}
    e= \frac{1}{M} - \frac{u}{2^k}
\end{equation}
where  $e$ needs to satisfy :
\begin{equation}
    M\cdot e <1
    \label{eq:barret_error}
\end{equation}
The value of $k$ was incremented until it satisfies the condition in Eq. (\ref{eq:barret_error}).

%% file: sections/3_proposed.tex
\section{ Proposed Hardware for Pipelined Polynomial Multiplier}
\label{sec:proposed}

This section has two subsection. First subsection explain the idea of pieplined NTT and next subsection details the hardware architecture.

\subsection{Pipelined NTT}
A pipelined design can reduce total number of clock cycles required to perform the NTT operations. Each stage can start performing butterfly operations as soon as the first pair of data in the correct order becomes available.

In general total number of clock cycles required to compute first NTT using  pipelined design could be calculated as:
\begin{align}
    &pip\_ntt\_num{\_}clocks \nonumber \\
    & = N/2(1+1/2+1/4+..+1/(2^{log^N_2}))+log^N_2-1 \nonumber \\
    &=N+log^N_2-2
    \label{tot_clks}
\end{align}

\begin{figure}
    \centering
    \includegraphics[scale=0.5]{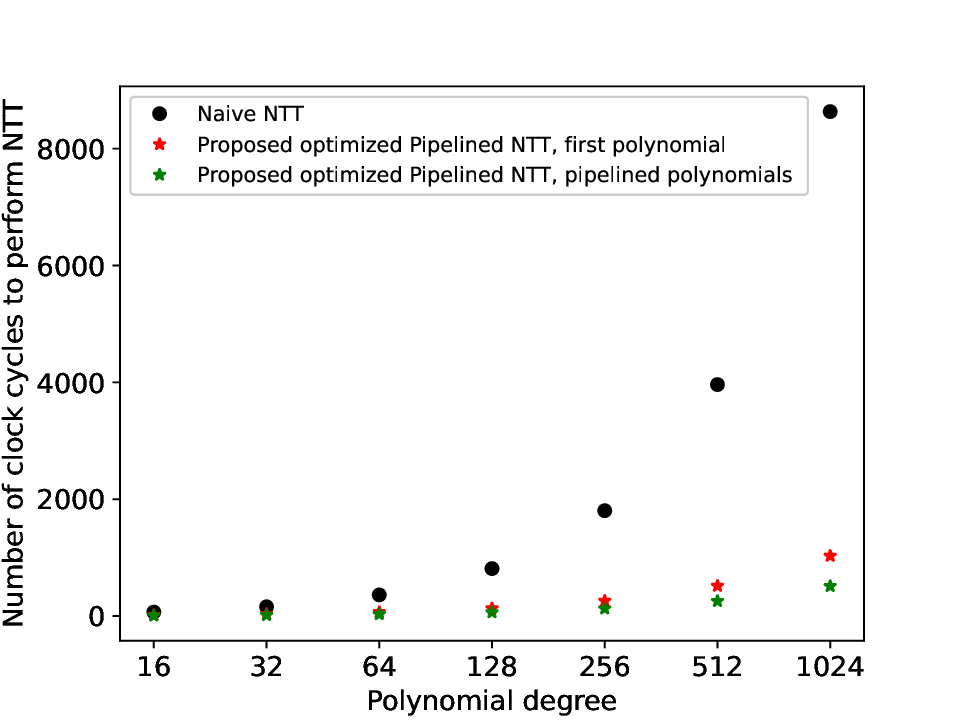} \\
    \caption{ Total number of clock cycles required for naive and proposed optimized pipelined implementation of NTT. In the case of  pipelined NTT, total number of clock cycles required to perform NTT on the first polynomial is higher as coefficients have to move through the pipeline. For the following polynomials, it only takes N/2 clock cycles to perform NTT on a degree n polynomial.
    (b) Total number of registers required to implement  naive and proposed optimized pipelined NTT.  }
    \label{fig:nor_pip_num_clks_regs}
\end{figure}

The Number of clocks to compute polynomial multiplication on the first two polynomial entering the pipeline can be calculated  as follows:
\begin{align}
    &mul\_pip\_num{\_}clocks \nonumber \\
    & = [1]+[(N+log^N_2-2) ]+[1]+[(N+log^N_2-2)] +[1] \nonumber \\
    &=2N+2log^N_2-1
\end{align}
It should be noted that this value is only for the first polynomials and later polynomial multiplication takes $N/2$ clocks to complete. Total number of clock cycles to compute  pipelined NTT and naive NTT is compared in the Fig. \ref{fig:nor_pip_num_clks_regs}. As displayed in Fig. \ref{fig:nor_pip_num_clks_regs}, total number of clock cycles to compute pipelined NTT is considerably smaller comparing to naive NTT.

\begin{figure}
    \centering
    \includegraphics[scale=0.3]{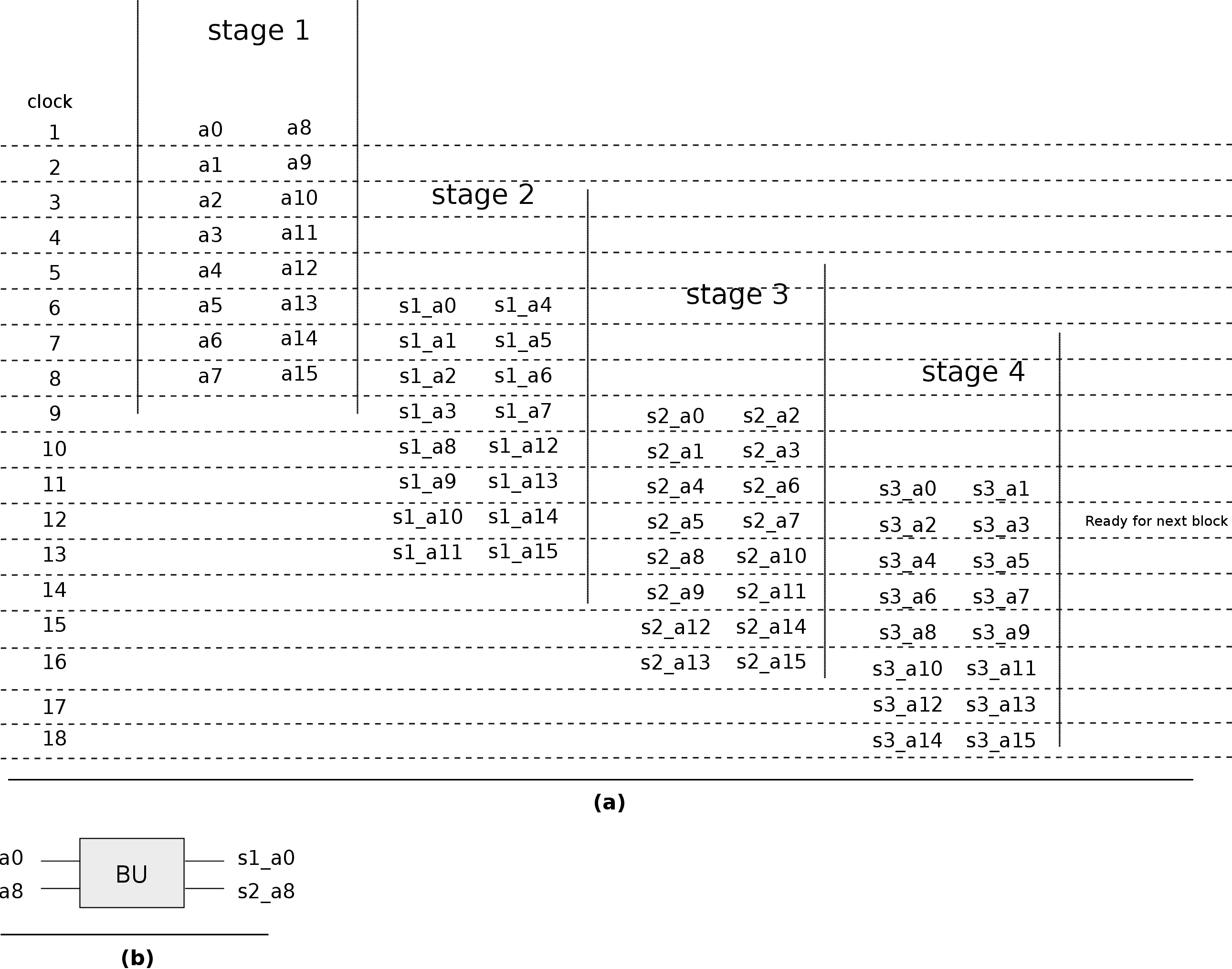}
    \caption{Pipelined computation in NTT. In each stage, butterfly operations can start as soon as the operands become available.
    At stage 1, butterfly operations start immediately as $a0$ and $a8$ in this stage are original polynomial coefficients and are available.
    It was assumed that result of butterfly operation on polynomial coefficient $ak$ at stage ($si$)  called $si\_ak$ as shown in (b) for the example of $a0$ and $a8$.
    Stage 2  starts when $s1\_a0$ and $s1\_a4$ are available which takes 5 clock cycles.   When fully pipelined, NTT takes 8 clock cycles to complete. }
    \label{fig:pipelined_ntt}
\end{figure}

For the case example, a polynomial with $N=16$, it takes $18$ clocks cycles to compute NTT on the first polynomial entering pipeline comparing to naive implementation in the previous section which required $32$ clock cycles. It should be noted that in the case of back-to-back polynomials, it would take only take $16$ clock cycles to compute NTT. As an example, the first butterfly operation in stage $2$ of Fig. \ref{fig:ntt_normal} (a) is on
$s1\_a0$ and $s1\_a4$ which are result of butterfly operations on  $a0$-$a8$ and $a4$-$a12$. The $s1\_a0$ and $s1\_a4$ would be available after cycle $5$.
Fig. \ref{fig:pipelined_ntt} describes how pipelining can be used to design optimized  pipelined NTT where butterfly operations in each stage starts as soon as the required operands become available.

In this work, to create the pipelined architecture, instead of memories, shift \ac{fifo}s are used. It is assumed that in each clock cycle, a pair of coefficients are send to pipelined NTT.

Each stage, requires to hold a specific number of coefficients stored in registers. As an example, in stage $2$ of the Fig. \ref{fig:pipelined_ntt}, ten  registers are required to hold coefficients for $5$ clock cycles until $s1\_a4$ becomes available.

In general, total number of registers required to hold the data for each NTT module could be calculated as:
\begin{equation}
    pip\_ntt\_num{\_}regs=n+2log^n_2-2
\end{equation}
The total number of registers for the NTT polynomial multiplier implemented as Fig. \ref{fig:structure} is:
\begin{equation}
    pip\_mul\_tot{\_}regs=3~pip\_ntt\_num{\_}regs+8
    \label{eq:pip_ntt_num_regs}
\end{equation}

For the case example, after generating $s1\_a0$ and $s1\_a4$, the corresponding registers are set free since these coefficients are no longer needed.  \ac{fifo} now shifts and available for new data received from  stage $1$. Similarly, stage $3$ needs $6$ registers, while stage $4$ needs $4$ and the last stage needs $2$ registers. The total number of registers required for calculating pipelined NTT is $18$ registers, which is the same as memory required to perform naive NTT in the Fig. \ref{fig:ntt_normal}.

The other resource to compare is number of butterfly units. The total number of clock cycles for the fully pipelined design to calculate first polynomial multiplication between two $N$-degree polynomials is as shown in Eq. \ref{eq:pip_ntt_num_regs}. However, the main advantage of the proposed approach is that after feeding the first polynomial, the second polynomial could be sent to the multiplier immediately. This way,
it takes only $N/2$ clock cycles to compute polynomial multiplications. In general, total number of butter fly units required for pipelined NTT is the same as number of stages and is equal to $log^{N}$.

Back to example case of the polynomial with degree of $16$, the pipelined NTT requires $4$ butterfly units as these unit can work simultaneously. 

In the next sub section, the hardware to perform polynomial multiplication is explained.

\subsection{Pipelined NTT Hardware Architecture}

In this section, the details of the hardware for pipelined multiplier is explained. This design is based on $N=256$ and $M=1,049,089$. The $M$ requires data path and register with width of $21$-bit.
\subsubsection{Step 1: Weighting polynomial coefficients}
In this step all values of $\phi^i ~mod ~M$  were calculated and stored in a \ac{rom}. When the multiplication starts, the reading address from ROM starts to increment to load the corresponding $\phi^i$ value for each coefficient.

\begin{figure}
    \centering
    \includegraphics[scale=0.52]{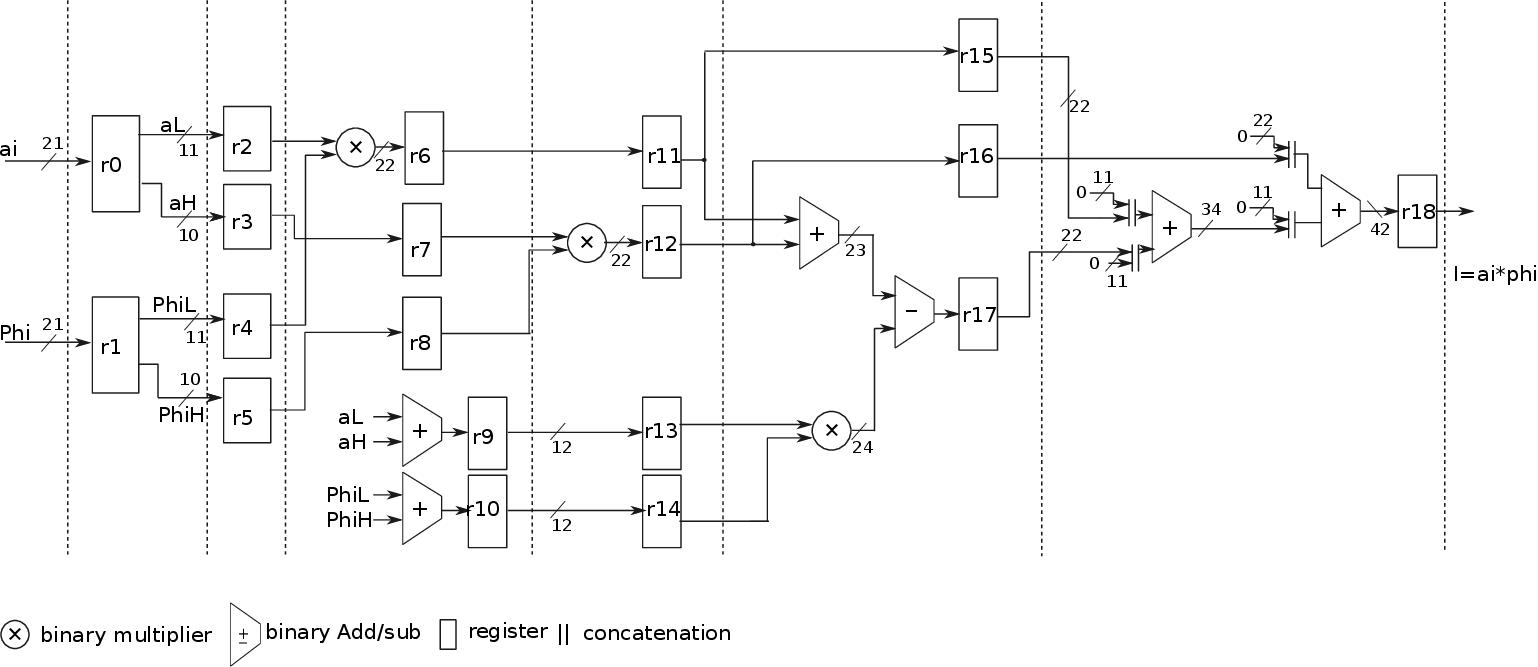}
    \caption{Splitting the binary multiplication of $\phi_i$ and $a_i$ using \ac{ka}  algorithm to increase speed and frequency. First, multiplicand are  captured in register $r_0$ and $r_1$. $\phi_i$ and $a_i$ then split into two parts of $L$,  $H$ and stores in registers $r2$ to $r5$. In next three clock cycles (each dash line represent a clock cycle) smaller sub-multiplications in \ac{ka} are performed. Last clock cycle construct the multiplication by shifting and adding sub-multiplications results.   }
    \label{fig:kmul}
\end{figure}

In the proposed design, objective is to reduce the critical path delay and increase the design frequency. Therefore, the \ac{ka}  algorithm was used to divide the $21$-bit binary multiplication into three smaller multiplications. The data flow graph for digital implementation of the multiplier using \ac{ka}  algorithm is shown in the Fig. \ref{fig:kmul}.

The multiplication is followed by a modulus reduction of $M$  using the Barret reduction algorithm. The reduction algorithm was optimized for hardware implementation for specific value of modulus and operand size.

\begin{figure}[htp]
    \centering
    \includegraphics[scale=0.6]{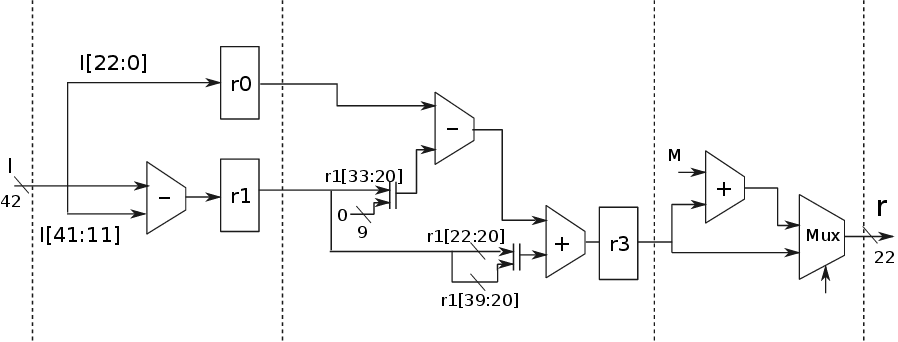}
    \caption{Implementation of reduction modulus $M$ using Barret algorithm. The hardware is optimized for $M=1049089$, $u=1048064$ and $k=40$.
    \label{fig:barret_red} }
\end{figure}

Minimum   values of $u$ and $k$ that satisfies the Eq.  (\ref{eq:barret_error}) are
$u=1,048,063$ and $k=40$.

However, in this design $u$ and $k$ was selected as $u=1,048,064$ and $k=40$ since   $1,048,064=2^{20}-2^{9}$ is one subtracter cheaper to implement comparing to  $1,048,063=2^{20}-2^9-1$.   $M$  could  be rewritten as $M= 2^{20}+2^9+1$. The Eq. (\ref{eq:barret_find_quotient}) and  (\ref{eq:barret_find_remainder}) can be computed as:
\begin{align}
    \label{eq:barret_red_shift_add}
    I_r&= 2^{-40}\cdot (2^{20}I-2^{9}I)  \cdot ( 2^{20}+2^9+1) \nonumber \\
    &=2^{-20} \cdot(I-2^{-11}I) \cdot   ( 2^{20}+2^9+1)
\end{align}
The optimized hardware is shown in Fig. \ref{fig:barret_red}. In this figure, $I-2^{-11}I$ is calculated by shift and add operation and result is stored in register $r1$.  Multiplication in  $2^{-20}$ is performed by selecting the $r1[41:20]$. The value of $I_r$ can be computed as follows: 
\begin{align}
    I_r=I&-r1[41:20]( 2^{20}+2^9+1)\nonumber \\
    =I&- r1[41:20]\nonumber \\ &- r1[41:20],9'b0 \nonumber \\
    &-r1[41:20],20'b0&
\end{align}
Therefore, it is  required to select the $23$ least significant bits, which is the size of $M$ plus a bit for addition/subtraction overflow. 
\begin{equation}
    \begin{split}
        I_r= I&- \\
        r1[41:20]&- \\
        r1[33:20],9'b0&-\\
        r1[22:20],20'b0&
    \end{split}
\end{equation}
$r[41] $ is the sign bit and is zero since $r1$ is result of multiplication of polynomial coefficients with twilling factors or $phi$. r[40] would also be zero as maximum value of $r1_max=(M-1)(M-1) - (M-1)(M-1) 2^(-11)      =XDF194744D7$. Therefore $r1[41:20]-r1[22:20],20'b0$ was replaced with concatenation ($r1[22:20],r1[39:20]$).

\begin{figure}
    \centering
    \includegraphics[scale=0.5]{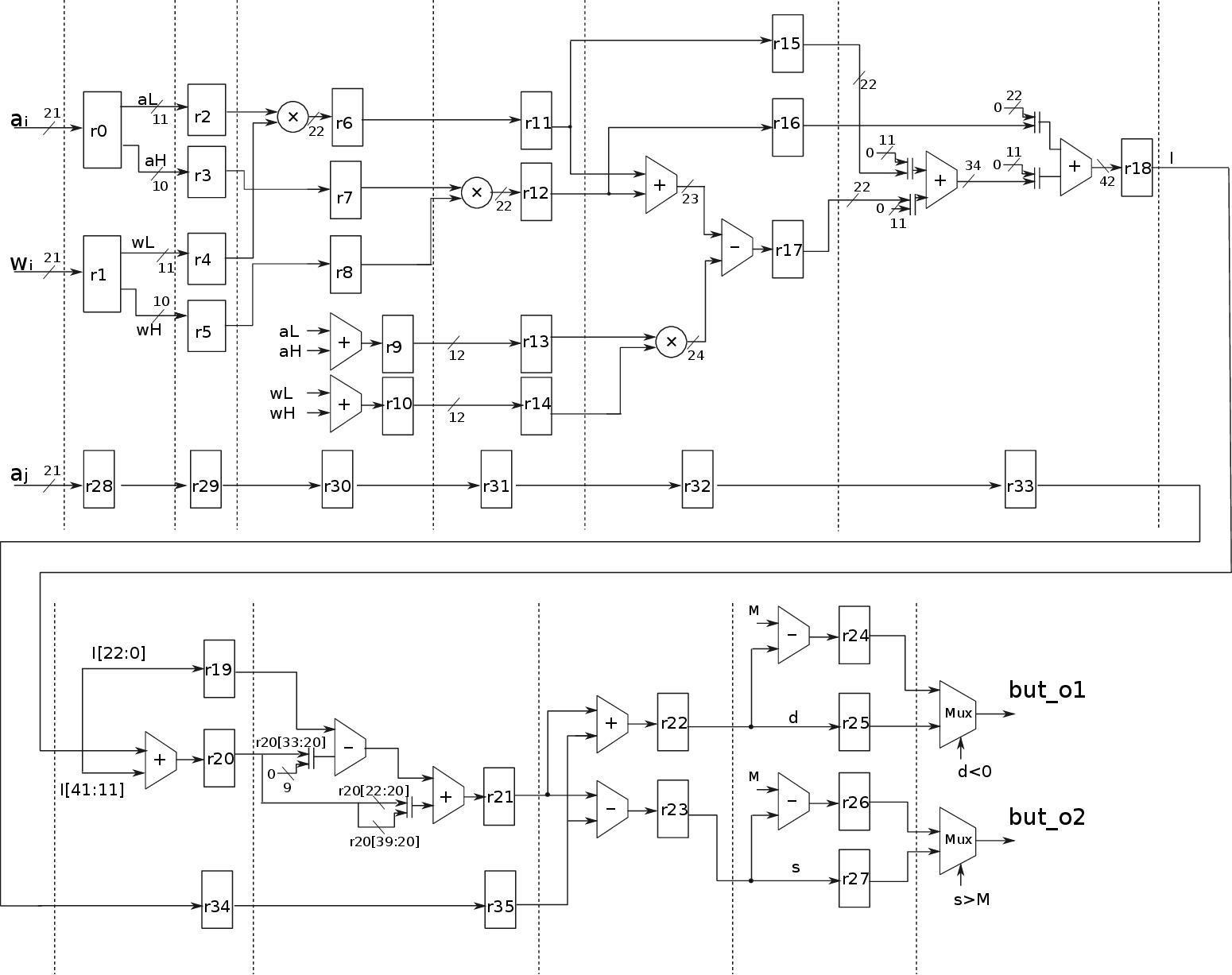}
    \caption{This figure shows the data flow graph for the butterfly unit used in this design. First, similar to Fig. \ref{fig:kmul}, coefficient $a_i$ of the polynomial $a$ is multiplied in the corresponding twiddling factor ($w_i$). The multiplication result is further reduced using the optimized Barrret reduction hardware described in Fig. \ref{fig:barret_red}. Meanwhile,  the other polynomial coefficient ($b_i$) is moved through pipelined registers. $b_i$ then added and subtracted from ($a_i$) and ($b_i$) multiplication result. In last clock cycle addition and subtraction results are compared against modulus $M$ for final butterfly results.   }
    \label{fig:butterfly}
\end{figure}

\subsubsection{ Step 2: NTT} 
In this step the NTT is performed on the polynomial coefficients. The NTT was performed in a similar method to Fig. \ref{fig:pipelined_ntt}. However the polynomials have a degree of  $n=256$, therefore, the NTT is performed in $8$
stages instead of $4$ for the case of $n=16$.

{\bf Butterfly Block} The hardware for the butterfly block is shown in the Fig. \ref{fig:butterfly}. In this figure,  first, polynomial coefficients ($a_i$) and their corresponding twiddling factors ($\omega_i$) are multiplied using \ac{ka} algorithm.     The Barret reduction circuit in Fig. \ref{fig:barret_red} is used to perform reduction modulus $M$ on the \ac{ka} multiplication result.

 The other polynomial coefficient input of butterfly unit ($aj$) is then  added  to output of Barret circuit and result is written to $r22$ register.  $aj$ is also subtracted from Barret circuit output and result is written to $r23$ register. Contents of these two registers are then compared against $M$ to perform final reductions and producing results of butterfly operation.
As depicted in Fig. \ref{fig:butterfly}, the butterfly hardware is fully pipelined. In each clock cycle, new coefficients and twiddling factors are received and processed by this unit.

{\bf NTT  Stages:} Each stage of NTT multiplier is described in the following.
\begin{itemize}
    \item {Stage 1:~}
        This stage receives the result of the previous step (multiplication in $\phi^i$) as inputs in the correct order, performs butterfly operation and send the data to next stage immediately. The twiddling factor (w) for this stage is always one. Therefore, butterfly operation can be easily performed with additions and subtractions. The hardware for this stage is depicted in Fig. \ref{fig:ntt:st1}.
        \begin{figure}
            \centering
            \includegraphics[scale=0.8]{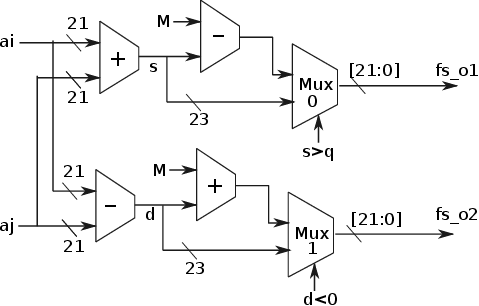}
            \caption{Stage 1 in the NTT module. The $\omega$=1 for this stage and therefore the polynomial multiplication can be replaced with addition and subtraction. }
            \label{fig:ntt:st1}
        \end{figure}
    \item{ Stage 2:~}
        In this stage, the  output from the previous stage need to be stored for 64 clock cycles before the butterfly operation in this stage could start.
        \begin{figure*}
            \centering
            \includegraphics[scale=0.65]{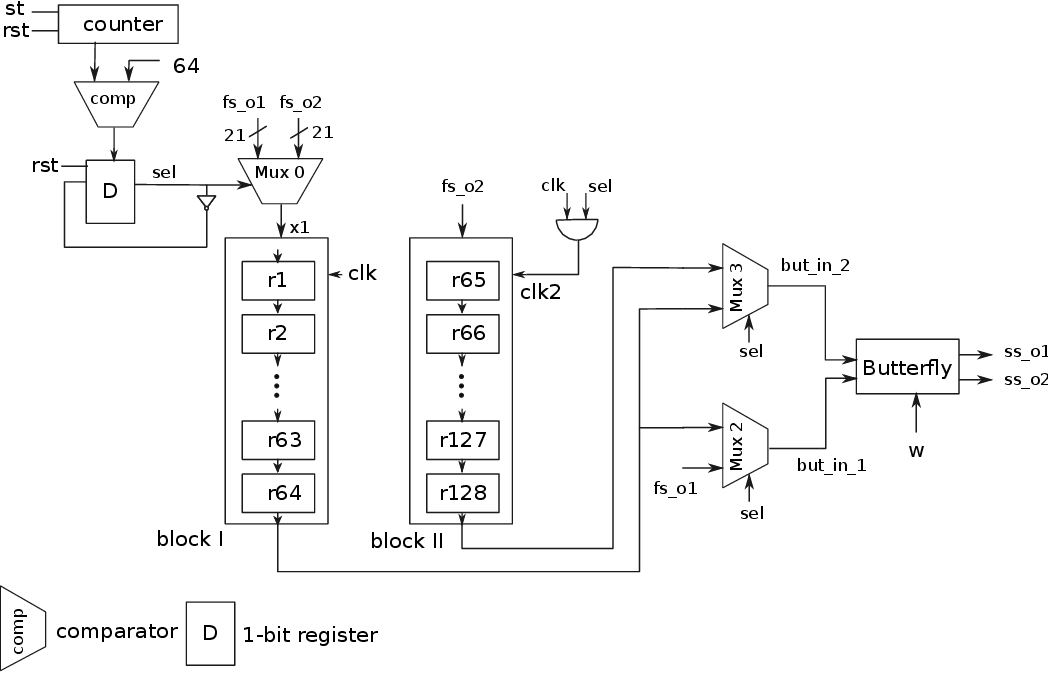} 
            \caption{Proposed FIFO based pipeline for the storing incoming polynomial coefficients. }
            \label{fig:ntt_st2}
        \end{figure*}
        This stage requires 128 registers to store data, which is the largest of all stages. The hardware to implement this stage is shown in Fig. \ref{fig:ntt_st2}.     Counter in this hardware checks $st$ signal to start processing data.
        The initial value of $se$ register is 1. At the beginning both block I and block II registers will be loaded with the data arrived from the previous stage. After 64 clock cycles, the first data is ready to be sent to butterfly unit. The value of $sel$ changes and contents of $r64$ and $fs\_o1$ (now arrived from previous stage) goes into the butterfly unit. The clock to block II is gated with $sel$ signal. Therefore, this block stops loading the new data. The new data ($fs\_o2$) will be loaded to block I instead. The other input ($fs\_o1$) and $r64$ are now consumed.

After another 64 clock cycles, the $sel$ changes again. The new data ($fs\_o2$) will be loaded into block II now and the output of this block ($r128$) and ($fs\_o2$) will be consumed by sending to butterfly unit.
        The truth table is shown in Table \ref{table:st2_table}.
        \begin{table}
            \centering
            \caption{Truth table for the hardware in Fig. \ref{fig:ntt_st2} (X: Don't care)}
            \label{table:st2_table}
            \centering
            \begin{tabular}{c|cccccc}
                counter&sel&clk2&x1&but$\_$in$\_$1& but$\_$in$\_$2\\ \hline
                $<64$ &   1&clk&ss$\_$o1&X& X\\ \hline
                $\geq 64~\& <128$ &   0&0&ss$\_$o2& ss$\_$o1&r64\\ \hline
                $\geq 128~\& <192$ &   1&clk&ss$\_$o1& r64&r128\\ \hline
            \end{tabular}
        \end{table}

        After 128 clock cycles, the $sel$ changes again and new data will be loaded into both block I and block II. This new data does not belong to the current polynomial since already all 256 polynomial coefficients are loaded or consumed. In the case of pipelined multiplier, these data are polynomial coefficients of the next polynomial. After 192 cycles, the counter  resets and checks the $st$ signal to start computation again.

        The value of $w$ in this stage changes between 4 values. Therefore, instead of saving it into a memory, we used registers and multiplexers to select the correct value of $\omega$.
    \item {Stage 3:~}
        Stage 4 is similar to stage 3. The difference is, in this stage the data received from the previous stage needs to be stored for 32 clock cycles. Therefore, this stage requires 64 registers to store data. The $sel$ signals changes every 32 clock cycles comparing to 64 in stage 3. The value of $w$ in this stage changes between 8 values which are saved in registers.
    \item {Stage 4, 5, 6, 7 and 8}
        These stages are similar to the previous stages of 3 and 4. The number of registers required for these stages to hold the data is smaller. For stage 4 is 16, for stage 5 is 8, for the stage 6 is 4, for the stage 7 is 2 and finally for the stage 8 is 1 registers. The difference between stage 4, 5, 6, 7 and 8 and previous stages  of 1, 2 and 3 is that the number of $w$ coefficients required to perform polynomial multiplication becomes larger. Therefore, we used a memory to store these values and load the $w$ into butterfly.
\end{itemize}
\subsubsection{Step 3: Element-Wise Multiplication}
In this step  the NTT result of the polynomial $a$ and $b$ are multiplied. The multiplier used in this stage is the same  multiplier depicted Fig. \ref{fig:kmul}.
 

\subsubsection{Step 4: Inverse NTT}
This step takes inverse NTT of the polynomial obtained from the element-wise multiplication.
The first inverse stage is similar to stage 2 of the NTT transform in the Fig. \ref{fig:ntt:st1}.
In stage 2 of inverse NTT, the data need to be stored for the 2 clock cycles. The number of clock cycles which data needs to be stored increases for the next  stages where inverse stages of 3, 4, 5, 6 and 7 hold the data for 4, 8, 16, 32 and 64 clock cycles.   Stage 8 is similar to stage 1 of NTT but with 128 different values of inverse $w$.
\subsubsection{Removing polynomial weights }
This is the final step of multiplication where values of $\phi^{(-i)}~mod~M$ are multiplied into the inverse NTT results. The hardware of this stage is the same as Fig. \ref{fig:kmul} and \ref{fig:barret_red} except that that it uses different value of $\phi$.

%% file: sections/4_results.tex
\section{Result and Discussion}
\label{sec:results}
As discussed in the previous sections, the proposed  pipelined NTT implementation requires considerably lower amount of clock cycles comparing to naive implementations of NTT (see Fig. \ref{fig:nor_pip_num_clks_regs}). The strength on the proposed design is that not only it is faster but it requires the roughly the same number of registers as  naive NTT implementation. The number of butterfly units however,  is higher for the pipelined NTT. NTT based polynomial multiplication requires computation of NTT and inverse NTT of the polynomials as shown in Fig. \ref{fig:structure}. Using a fast and efficient NTT hardware has a great impact on the speed and efficieny of the multiplier.

\begin{table}
    \centering
    \small
    \caption{Comparison of FPGA resource utilization and speed for polynomial multiplication.}
    \label{comparison}
    \begin{tabular}{ccccccc} \hline
        \multirow{2}{*}{Design} & \multirow{2}{*}{Slice} & \multirow{2}{*}{DSP} & \multirow{2}{*}{BRAM} & \multirow{2}{*}{Cycles} & \multirow{2}{*}{Freq. (MHz)} & \multirow{2}{*}{ Delay ($\mu s$)}  \\
         &  &  &  &  &  &     \\ \hline
        \cite{ntt_chen} & 886 & 4 & 4 & 1618 & 258 & 6.27  \\ \hline
        \cite{nttdu} & 953 & 16 & 4.5 & 917 & 247 & 3.72   \\ \hline
        \cite{ntt_feng} & 14k & 128 & 1 & 220 & 235 & 0.94  \\ \hline
        This work& 8093 & 116 & 9 & 128 & 234 & 0.56  \\ \hline  \hline
            \multirow{2}{*}{Design} & \multirow{2}{*}{q} & \multirow{2}{*}{n} & \multicolumn{2}{c}{\multirow{2}{*}{FPGA}} & \multirow{2}{*}{} & \multirow{2}{*}{}  \\
         &  & & \multicolumn{2}{c}{\multirow{2}{*}{}}   &  &       \\ \hline
        \cite{ntt_chen} & 1,049,089&256 &  \multicolumn{2}{c}{ Spartan-6 XC6SLX100 -3 } &  &     \\ \hline
        \cite{nttdu} &1,049,089 & 256 & \multicolumn{2}{c}{ Spartan-6 LX100 }  &  &      \\ \hline
        \cite{ntt_feng} &1,049,0890& 256 & \multicolumn{2}{c}{ Spartan-6 XC6SLX150 }  &  &     \\ \hline
        This work&  1,049,0890& 256 & \multicolumn{2}{c}{ Spartan-6 XC6SLX150 }  &  &      \\ \hline
    \end{tabular}
\end{table} 

Further, the proposed NTT polynomial multiplier were synthesized using Vivado  Design Suite and Implemented on FPGA as a proof of concept.
The FPGA resource utilization and speed is compared with similar works in Table \ref{comparison}. In a fully pipelined state, this multiplier is capable of completing polynomial multiplication between two polynomials with degree $n=256$ and modulus $M=1049089$ (21-bit) in 128 clock cycle.
This number of clock cycles is the smallest reported yet. Nevertheless, it utilizes less slices and DSP blocks comparing the fastest NTT multiplier implementation reported in literature.
The proposed design runs in a frequency close to similar works. The total delay for this design is 0.56 $\mu s$.

%% file: sections/5_conclusion.tex
\section{Conclusion}
\label{sec:results}
In this work a FIFO based circuit was presented for a fast and efficient calculation of large polynomial multiplication using number theoretic transformation. The design was implemented on FPGA and results were compared to similar works. The proposed design is capable of calculating polynomial multiplication of two polynomials with degree of $n$ in $n/2$ clock cycles. The implementation results also indicated that the proposed design uses less resources compared to the fastest similar work presented to calculate polynomial multiplication.